\documentclass{iau}
\usepackage{graphicx}

\title[The data sharing advantage] 
{The data sharing advantage in astrophysics}

\author[Bertil F. Dorch, Thea M. Drachen \& Ole Ellegaard]   
{Bertil F. Dorch, Thea M. Drachen \and Ole Ellegaard}

\affiliation{
University Library of Southern Denmark, 
University of Southern Denmark, \\ Campusvej 55, DK-5230, Odense, Denmark 
  }

\pubyear{2015}
\volume{xxx}  
\pagerange{xxx--xxx}
\jname{FM 3: Scholarly Publication in Astronomy}
\editors{Marsha Bishop, Eds.}
\begin{document}

\maketitle

\begin{abstract}
We present here evidence for the existence of a citation advantage within astrophysics for papers that link to data. Using simple measures based on publication data from NASA Astrophysics Data System we find a citation advantage for papers with links to data receiving on the average significantly more citations per paper than papers without links to data. Furthermore, using INSPEC and Web of Science databases we investigate whether either papers of an experimental or theoretical nature display different citation behavior.
\keywords{astronomical data bases, methods, sociology of astronomy, statistical}
\end{abstract}

\firstsection 

\section{Introduction}

Research funders increasingly require data management plans prior to applications for funding. Similarly, infrastructures and policies are arising regarding both archiving, documentation and sharing research data. While scientists are increasingly being evaluated and funded according to quantitative measures, e.g. by citations, it is relevant to ask whether a citation advantage exists that is related to the activity of sharing data, similar to the debated citation advantage related to Open Access (e.g. \cite[Kurtz et al. 2005]{Kurtz+ea05} and \cite[Kurtz et al. 2007]{Kurtz+ea07}).

We present here a simple study of astrophysical publications to investigate a possible increased citation impact resulting from linking to data, using the NASA Astrophysics Data System, henceforth ADS (cf. \cite[Kurtz et al. 2000]{Kurtz+ea00}). This work is an extension of the initial study concerning data links in papers published in the journal {\em ApJ} during 2000--2010, presented in an unpublished working paper by \cite[Dorch (2012)]{Dorch+ea2012}. 

\section{Publication data and method}

The ADS, launched by NASA in 1992, is hosted by the Harvard-Smithsonian Center for Astrophysics. ADS is an online publication database of millions of astronomy and physics papers receiving abstracts or tables of contents from  hundreds of journal sources. The ADS also lists citations for each paper. The ADS search engine is tailor-made for searching astronomical abstracts and can be queried for author names, astronomical object names, title words, abstract text, and results can be filtered according to a number of criteria (cf. \cite[Eichhorn et al. 2000]{Eichhorn+ea00}). 

For each publication record in ADS, a number of links are possible, including data links to online data, e.g. at external data centers. Links of this type are abbreviated ``D'' aka. D links (cf.\ \cite[Accomazzi \& Eichhorn 2004]{Accomazzi+Eichhorn2004} and \cite[Eichhorn et al. 2007]{Eichhorn+ea07}). Therefore, it is possible to limit ADS to publications with or without D links. 

In part of the work presented here, we invoke also a secondary source of publication data, the INSPEC database from the Institute of Engineering and Technology (formerly the IEE) and a source of citation data, and the Web of Science (WoS) science citation index from Thomson Reuters. Like WoS, but unlike ADS, INSPEC is a commercial major indexing database of scientific and technical literature.

~\\
In this study, we perform two analyses:
\begin{enumerate}
	\item We investigate the number of papers and citations for papers with or without D links during the period 2000--2014 for {\em ApJ}, {\em A\&A} and {\em MNRAS} using NASA ADS.
    \item We investigate the number of papers and citations for experimental and theoretical papers respectively during 2010 for {\em ApJ}, {\em A\&A} and {\em AJ} using INSPEC and NASA ADS.
\end{enumerate}

Firstly, {\em (a)} we limit the study to papers published in major astrophysical journals during the 15-year period in the current millennium 2000--2014, cf.\ Table \ref{tab1}.
Furthermore, we define the citation advantage of papers that link to data $ P_{\rm D}$ as the ratio of the number of citations per year to papers with links to data, and the number of citations per year to papers without such links. Publication data and derivatives for {\em ApJ} are illustrated in Fig.\, \ref{fig1} left and right.

Secondly, {\em (b)} it is relevant to investigate whether we introduce a bias in selecting articles with data-links, e.g. whether experimental papers more often link to data, and whether experimental papers are cited more than theoretical papers.

~\\
To test this possibility, we apply the feature \textit{treatment type} that the database INSPEC assigns to all indexed papers:

\begin{itemize}
	\item \textit{Theoretical or mathematical} is assigned when the subject matter is generally of a theoretical or mathematical nature.
	\item \textit{Experimental} is used for documents describing an experimental method, observation or result. Includes apparatus for use in experimental work and calculations on experimental results.
\end{itemize}

Articles from the three journals {\em ApJ}, {\em A\&A} and {\em AJ} are downloaded into the reference handling program Endnote in order to extract DOIs for further processing. The relevant DOIs are then entered into INSPEC and the articles are separated into the two tiers: either classified as theoretical or experimental work. The few articles classified as both experimental and theoretical are discarded from the analysis. Finally, we apply, in this case, WoS in order to extract the number of citations because DOIs are not searchable in ADS.

{\em ApJ} as registered by ADS includes letters as well as the supplement series but the articles published in those latter categories are not fully included in WoS and we discard them from the present analysis. The number of articles with or without data links (as well as citation data from WoS) is then downloaded directly from ADS.

\begin{table}[htb]
  \begin{center}
  \caption{Data for the four journals {\em ApJ}, {\em A\&A}, {\em MNRAS} and {\em AJ}: Journal Impact Factor 2013 (JF), the average number of papers published per year during 2000--2014, $\langle N \rangle^{\rm papers}$, the average fraction of papers with D links $\langle n \rangle^{\rm papers}_{\rm D}$, the average fraction of citations resulting from papers with D links $\langle n \rangle^{\rm cite}_{\rm D}$, and the average D link citation advantage $\langle P_{\rm D} \rangle$ during 2000--2014.}
  \label{tab1}
 {\scriptsize
  \begin{tabular}{l|c|c|c|c|c}\hline 
{\bf Journal} & {\bf JIF 2013} & $\langle N \rangle^{\rm papers}$  & $\langle n \rangle^{\rm papers}_{\rm D}$ & $\langle n \rangle^{\rm cite}_{\rm D}$ & $\langle P_{\rm D} \rangle$ \\ \hline
{\em ApJ} incl.\ {\em ApJL} and {\em ApJS} & 6.280 & 3137 & 0.303 & 0.358 & 1.286 \\ \hline

{\em A\&A} & 4.479 & 1941 & 0.386 & 0.441 & 1.302 \\ \hline

{\em AJ} & 4.052 & - & - & - & 409 \\ \hline

{\em MNRAS} & 5.226 & 1725 & 0.239 & 0.247 & 1.055 \\ \hline

  \end{tabular}
  }
 \end{center}
\end{table}

A statistical analysis was performed as appropriate to test for any significance in mean citation counts between articles with and without datalinks as well as between theoretical and experimental articles. F tests were used to test for equal variance; two tailed t-tests were then run for unequal and equal variances as appropriate to test for significant difference between mean total citations per paper. Our focus is only on articles published in 2010. This ensures time to accumulate a sufficiently large number of citations.

\section{Results and discussion}

\begin{figure*}[htb]
\begin{center}
 \includegraphics[width=2.6in]{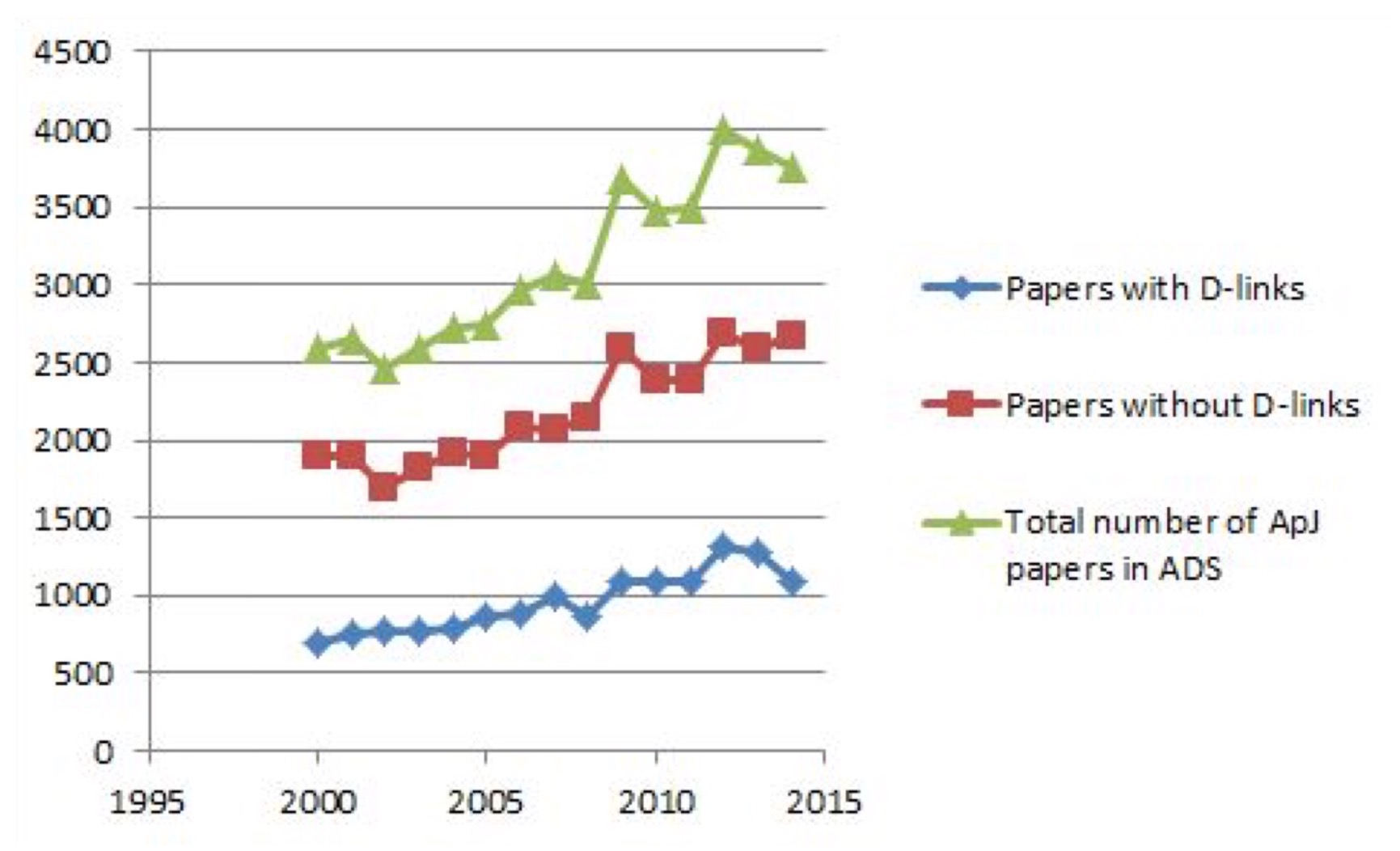} 
 \includegraphics[width=2.6in]{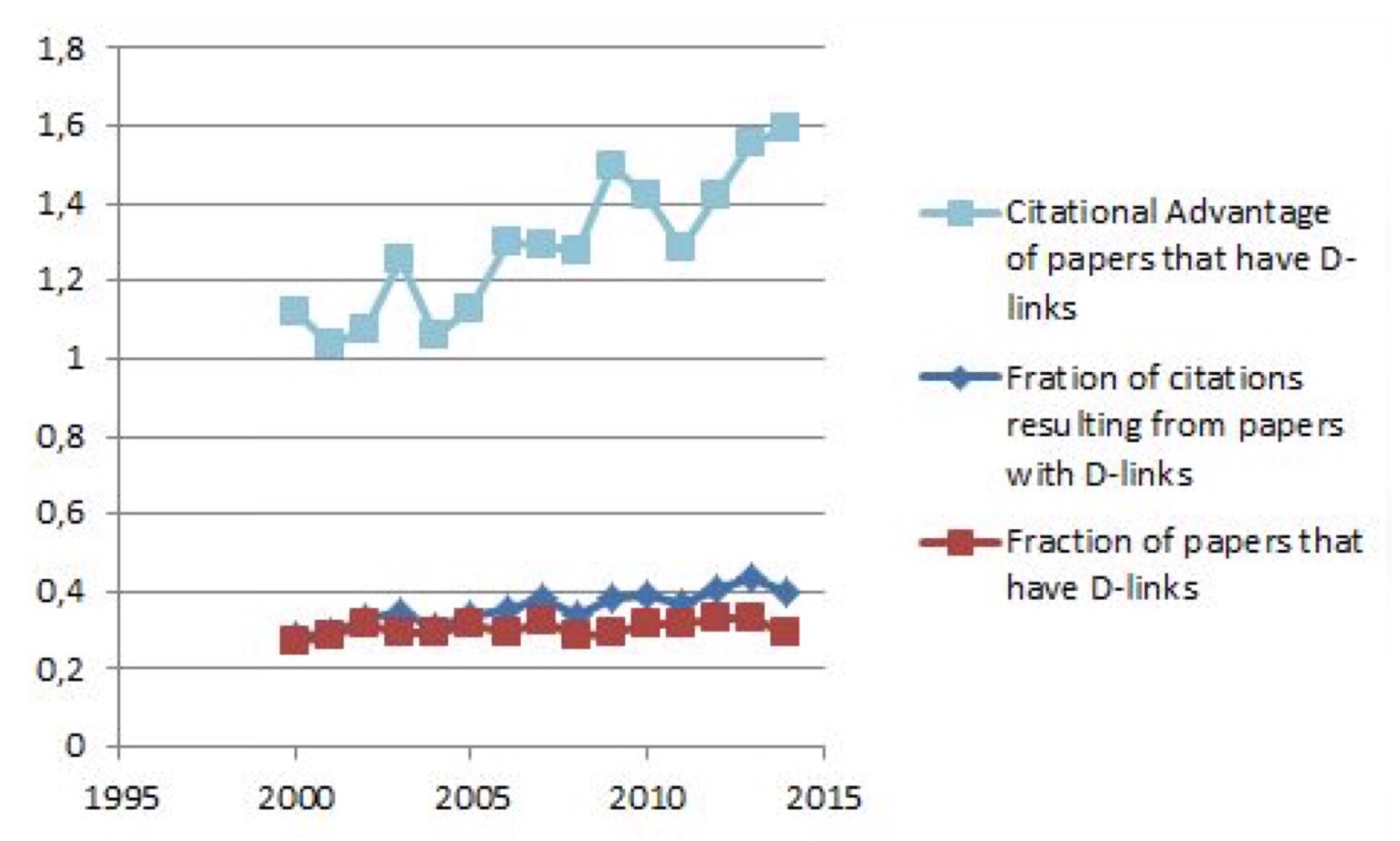} 
 \caption{Left: The number of papers in {\em ApJ} 2000--2014 as a function of the year of publication as registered in ADS. Upper curve (green): Total number of papers. Middle curve (red): Papers without links to data. Lower curve (blue): Papers with links to data. Right: Upper curve (blue): The citation advantage $P_{\rm D}$ as a function of the year of publication as registered in ADS. Middle curve (blue): The fraction of the total number of citations that result from papers with links to data  $n^{\rm cite}_{\rm D}$. Lower curve (red): The fraction of papers that actually have links to data $n^{\rm papers}_{\rm D}$.}
   \label{fig1}
\end{center}
\end{figure*}

The papers with D links received, in total, fewer citations per year on average relative to the papers without D links (by approximately a factor of two). However, there being fewer papers with links to data, it turns out that these papers on the average received more citations per paper i.e. during the examined period the D link papers in {\em ApJ} on average receive 28\% more citations per paper per year, than the papers without D links. Since 2009 that fraction is higher and in the case of {\em ApJ} more like 50\% more citations, cf. Fig.\,\ref{fig1} (right).

Next, we look at the journals and papers in term of their experimental or theoretical content. In case of papers published in {\em ApJ} the number of experimental papers is only slightly above the number of theoretical ones. The difference between the mean numbers of citations obtained by the two groups is small as well.

\begin{figure}[htb]
\begin{center}
 \includegraphics[width=2.6in]{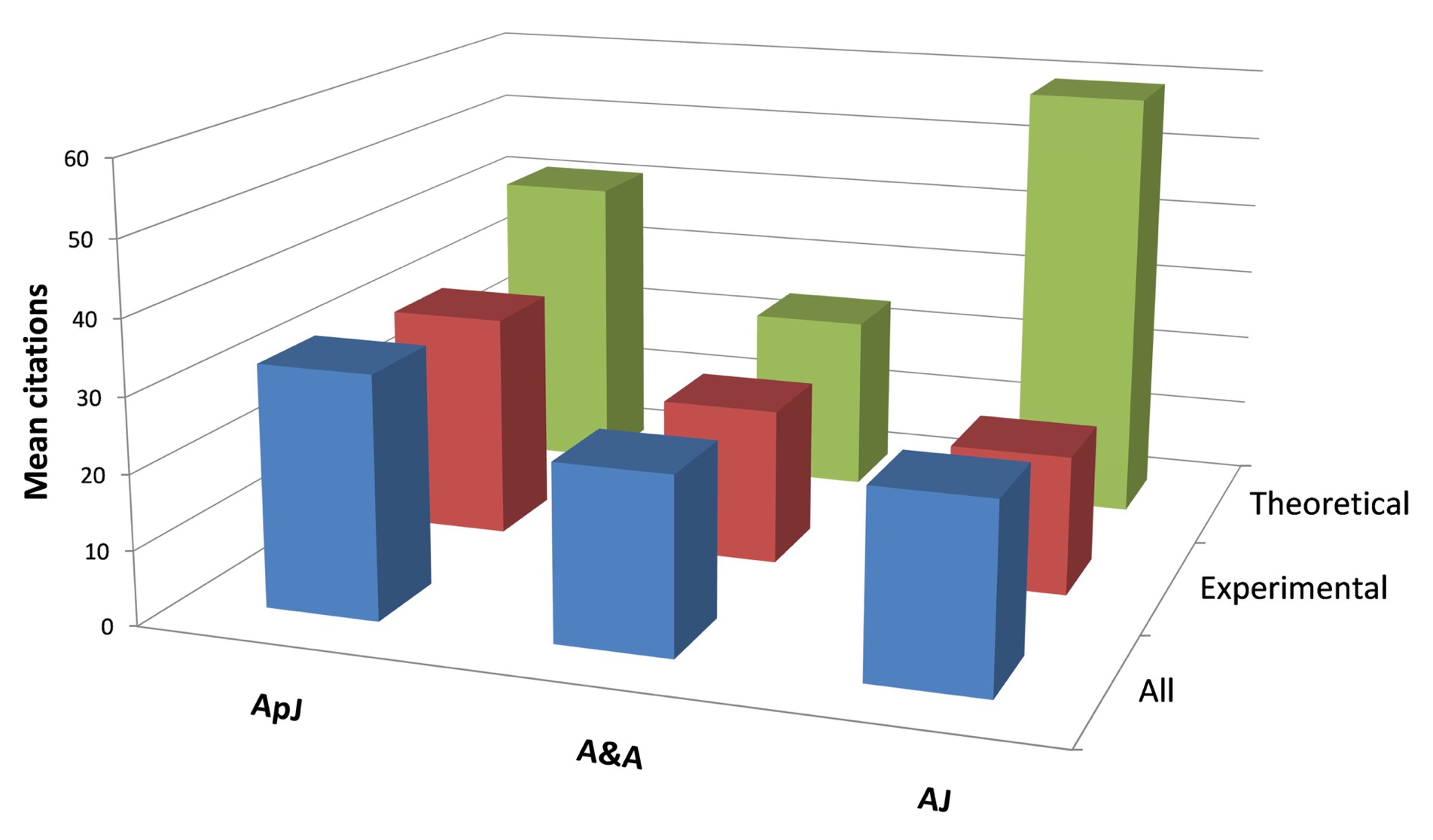} 
 \includegraphics[width=2.6in]{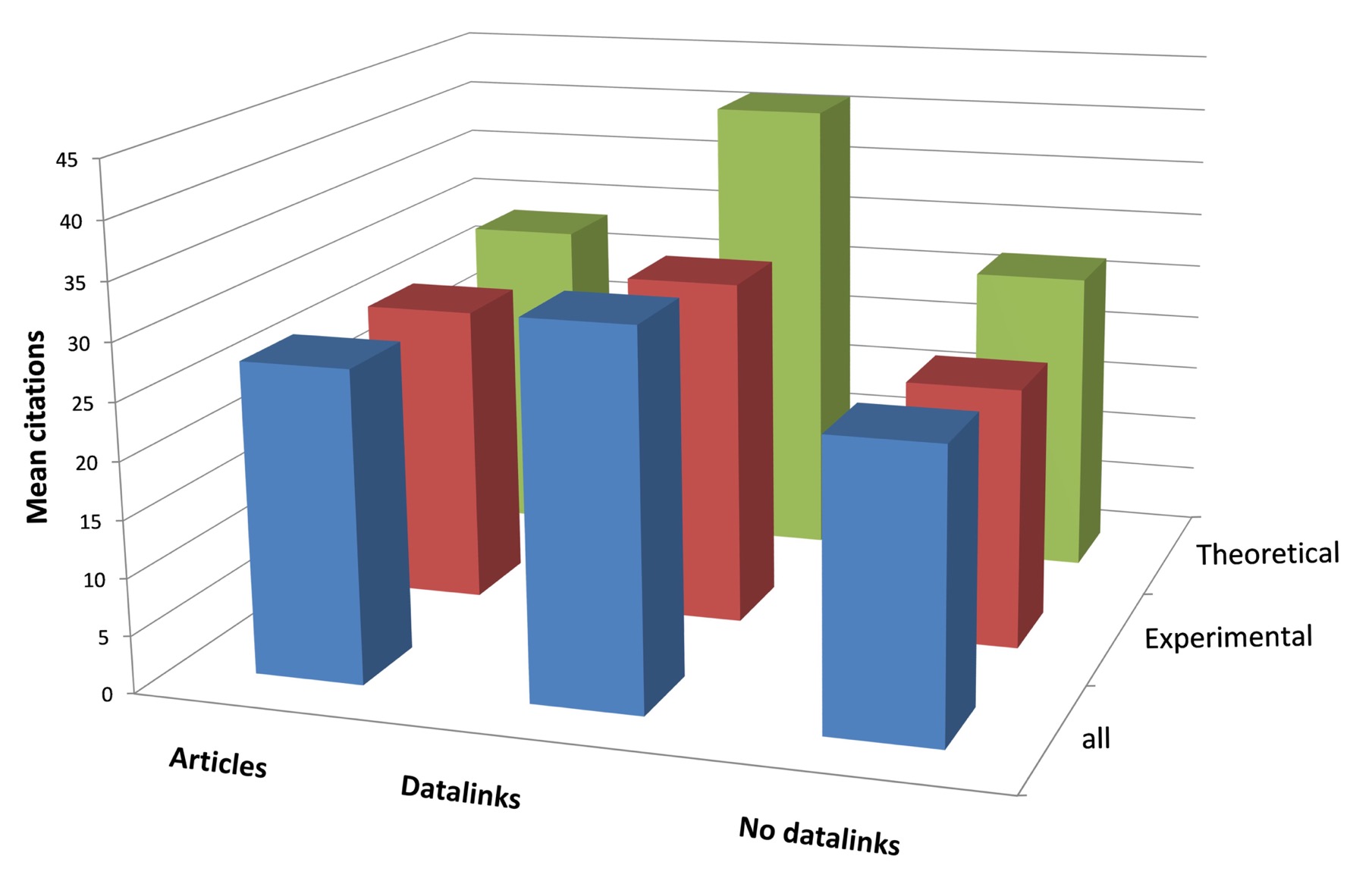} 
 \caption{Left: Histogram of the total mean number of citations for {\em ApJ}, {\em A\&A} and {\em AJ} papers with D links in 2010 and the corresponding contributions from experimental and theoretical papers. 
Right: Histogram of the total mean number of citations for {\em ApJ} in 2010 and for papers with and without D links (blue columns), and the differentiation for experimental papers (red columns) and theoretical papers (green columns).
}
   \label{fig2}
\end{center}
\end{figure}

The situation is different when considering papers with or without data links. In case of D link papers, the number of experimental papers is much larger than the number of theoretical papers, while the latter has the largest mean number of citations. On the other hand, the number of theoretical non-link papers is above the similar number of experimental papers, but still the theoretical articles obtain the most citations.

The same pattern is observed in case of the two other journals {\em A\&A} and {\em AJ}. The theoretical papers with data links obtain the highest number of citations. The difference is most pronounced in case of papers published in {\em AJ}, but this conclusion is based on rather few papers in the data. We have examined the statistical confidence level of our
conclusions: In case of {\em ApJ} and {\em A\&A} it is evident, although only evident at the 5\% significance level in case of {\em ApJ} ($p < 0.05$), that papers with D links obtain the largest numbers of citations. In case of {\em AJ} a $p$ value well above $0.05$ indicates that the citation advantage is not statistically well founded. In a similar fashion a significant advantage for obtaining citations has been observed for theoretical D link papers compared to experimental D link papers in case of all three journals. On the other hand, it can only be proven at the $p>0.05$ confidence level partly due to a low number of papers and scatter in citations data.

Our simple study indicates a clear tendency for papers with links to data to receive more citations per year on average, than papers that do not link to data. However, there are several biases that could be studied further, e.g.\ whether longer papers, papers with more authors etc.\ display generically different citation patterns. Also of potential importance is whether some subjects that ``naturally'' link to data have a higher citation impact than other fields, e.g.\ papers based on space missions or telescope data. 

\cite[Henneken \& Accomazzi (2011)]{Henneken+Accomazzi2011} performed an analysis restricting publication data using a set of 50 keywords: looking at cumulative citations to papers after a 10 year period. The report demonstrated a 20\% increase in citation count for papers with D links, compared to those without. Alas, evidence is mounting that linking to data enabling sharing does indeed merit those who do so. This evidence thereby also supports initiatives furthering the development of data infrastructure.

A more comprehensive account of the study presented in these proceedings will be published by \cite[Dorch et al.\ (2016)]{Dorch+ea2016}.

\acknowledgements

This research has made use of NASA's Astrophysics Data System Bibliographic Services.

\end{document}